# FLOW RATE ESTIMATION FROM PROBE VEHICLE DATA AND SAMPLE SIZE REQUIREMENTS

Mecit Cetin (Corresponding Author)
Assistant Professor
Department of Civil and Environmental Engineering
University of South Carolina
300 South Main St., Columbia, SC 29208 USA
+1-803-777-8952, cetin@engr.sc.edu

Gurcan Comert
Graduate Assistant
Department of Civil and Environmental Engineering
University of South Carolina
300 South Main St., Columbia, SC 29208 USA
+1-803-777-0665, comert@engr.sc.edu

## ABSTRACT

The interest to use probe vehicles for traffic monitoring is growing. This paper is focused on the estimation of flow rate from probe vehicle data and the evaluation of sample size requirements. Three cases are considered depending on the available information on the percentage of probes $p$ and the flow rate $\lambda$: i) $p$ is known but $\lambda$ is unknown ii) $\lambda$ is known but $p$ is unknown, and iii) both parameters are not known. Estimation methods for all three cases are presented along with the reliability of these estimates. For the first two cases, count data provide sufficient information to estimate the unknown parameters $p$ and $\lambda$ individually when the other one is known. For these two cases, simple analytical expressions are derived to analyze the accuracy and sample size requirements. However, when both parameters are unknown then additional information beyond count data is required. The position of probe vehicles in a queue at signalized intersections is used as additional information in that case. The results show for how long data need to be collected to estimate the parameters at acceptable confidence levels.

## INTRODUCTION

There is a growing interest to capitalize on the probe vehicle data for traffic control and management applications. Probe vehicles equipped with tracking and wireless communication technologies provide valuable information on the traffic conditions. With the recent Vehicle Infrastructure Integration (VII) initiative of the USDOT (*1*) the impetus to explore the use of probe vehicle data for ITS applications will become more prevalent. There are several key parameters such as flow, density, speed, travel time that are commonly used to describe different attributes of traffic flow or system state. In general, a probe vehicle can provide speed, location, and time data sampled from the near continuous time-space trajectory of the vehicle (i.e., GPS tracks). These data elements can be used to estimate the key traffic flow parameters mentioned above. Clearly, the market penetration level (i.e., the percent of vehicles serving as probes in the traffic stream) is a critical factor that impacts the ability to estimate system state reliably. Most of the existing studies on probe vehicle applications explore the relationships between the market penetration and the reliability of the travel time estimates (*2-4*). Network coverage is also an important issue that is addressed in the literature (*5-7*). Due to the complexity of the problem, none of these studies develop analytical models or closed form solutions that relate the number of probes to the reliability of the estimates. Instead, empirical analyses are performed in these studies that require data to be generated for numerous scenarios with different probe vehicle percentages. Typically, data from microscopic traffic simulation models are used for that purpose since real-world data with a large number of probes to support such analyses are not available.

This paper is focused on exploring the flow rate estimation from probe vehicle data. Flow rate is a key traffic flow parameter that is often used to describe the traffic demand. For example, if hourly traffic flow rate is known for each hour of the day one can estimate the ADT (average daily traffic) and AADT (annual average daily traffic). Traffic engineers typically install pneumatic tubes on local streets to estimate ADTs and AADTs if permanent count stations are not available on the links. Given that there could be potentially thousands of links in a large-scale urban network, estimating ADTs and AADTs on these links would require substantial resources. However, data from probe vehicles can be tapped into to estimate these quantities more effectively provided that there are sufficient probe vehicles throughout the network.

In order to estimate ADTs or AADT one can make certain assumptions that flow rate on a given day (e.g., Tuesday) and hour (or any other arbitrary time period) is constant. Then, probe observations corresponding to these particular time periods collected over multiple days or time periods can be treated as random samples which can be used to estimate the unknown parameter(s). The fundamental question is to determine how many samples are needed to reliably estimate the unknown parameter(s). The basic unknown parameter in this case is the actual flow rate (the average or mean of the flow rate). However, the percentage of probe vehicles on every link in the network may not be known reliably. Therefore, another parameter of interest is the actual percentage of probe vehicles.

The main objective of this paper is to investigate the impact of probe sample size or number of probe observations on the accuracy of the estimated flow rate. As mentioned above, the actual probe vehicle penetration level may not be known in real-world settings. Therefore, the paper also addresses the estimation of unknown market penetration level along with the flow rate. In this work, vehicle arrivals (or counts in a time period) are assumed to follow a Poisson process to simplify the analyses and to allow a comprehensive evaluation of the interactions of various parameters as explained in the next section. The analyses in this paper are geared more towards flow rates on signalized streets but some of the results are equally applicable to freeways.

**METHODOLOGY**

In probe-based traffic monitoring systems, a significant issue is the effect of market penetration level of probe vehicles (denoted hereon by *p*) on the ability to estimate traffic parameters accurately. Obviously, the higher the *p* the more accurate the estimates are. However, to aid decision making, relationships need to be established between *p* and the accuracy in order to determine what level of *p* would produce acceptance results. This often times is expressed in a classical confidence interval framework with specified confidence level and precision or deviation from the true mean. For establishing a relationship between *p* and accuracy of the estimated flow rate, overall the following parameters need to be considered:
- $\alpha$-level for the confidence level
- $\delta$: percent deviation from true value of the parameter being estimated. This is used to define the confidence interval.
- $\lambda$: arrival rate for all vehicles
- $\Delta$: observation period
- $p$: percentage of vehicles serving as probes in the traffic stream

As mentioned in the previous section, the arrival process is assumed to be a Poisson process with a constant rate ($\lambda$) in this paper. Clearly, traffic flow rate is not typically constant and exhibits a time-dependent behavior in general (tough for shorter time periods, arrival rate can be approximated to be constant). However, this assumption is made to reduce the number of variables and the complexity of the problem so that the interrelationships of the five parameters listed above can be investigated more thoroughly. Flow rate estimation in a dynamic setting is a topic of future research.

The parameter $\Delta$ listed above specifies the length of the interval over which probe vehicle data are collected. If the vehicle arrivals follow a Poisson distribution then, the probability function for the total number of vehicles ($N$) in this observation interval $\Delta$ (including probes) is as follows.

$$P(N=n|\lambda,\Delta) = \frac{e^{-\lambda\Delta}(\lambda\Delta)^n}{n!}, \lambda > 0, n = 0,1,2... \quad (1)$$

It is assumed that probe vehicles are randomly distributed within the traffic stream. In other words, every vehicle has the same likelihood of being a probe. Therefore, the relationship between the total number of probe vehicles ($N_p$) and all vehicles in interval $\Delta$ can be written as follows.

$$N_p = \sum_{i=1}^{N} y_i \quad \text{where } y_i \in \{0,1\}, \text{ and } P(y_i = 1) = p \quad (2)$$

The equation above asserts that every vehicle has equal probability of being a probe vehicle. Both $N$ and $N_p$ are two discrete random variables that follow Poisson distribution with rates $\lambda$ and $\lambda p$ respectively.

Given the arrival process as specified above, the flow rate ($\lambda$) and percent of probe vehicles ($p$) need to be estimated from the data of probe vehicles collected over an observation period of $\Delta$. Three scenarios can be conceived depending on the available information on $p$ and $\lambda$: i) $p$ is known but $\lambda$ is unknown ii) $\lambda$ is known but $p$ is unknown, and iii) both parameters are not known. This paper analyzes all these three cases.

For the first and second cases, the probe counts ($N_p$) collected over $\Delta$ provide the necessary information for parameter estimation. For estimating $\lambda$, the best predictor in the maximum likelihood sense is as follows.

$$\hat{\lambda} = \frac{N_p}{\Delta p} \quad (3)$$

Likewise, for estimating $p$ when $\lambda$ is known the following expression is used:

$$\hat{p} = \frac{N_p}{\Delta \lambda} \qquad (4)$$

Since the probability distribution of $N_p$ is known, the statistical properties of these two estimators ($\hat{\lambda}$ and $\hat{p}$) can be readily obtained. Therefore, the analyses can be performed without using simulation. The estimation of $\lambda$ and $p$ from probe counts is explained in the next two sections.

For the third case, when both $\lambda$ and $p$ are unknown, the probe count data by itself would not be sufficient to estimate both unknowns since the two are interdependent as shown in Equations 3 and 4, i.e., additional information is needed. The approach taken in this paper to obtain this additional information involves utilizing the location data of probe vehicles when they stop at signalized intersections. If a probe vehicle stops in a queue and transmits its location (e.g., distance from the stop bar), it might be possible to determine how many vehicles are ahead of this probe vehicle by assuming an average vehicle length (e.g., dividing the distance by the average vehicle length to find the number of vehicles). This additional information can then be utilized to estimate both $\lambda$ and $p$. In particular, the last probe vehicle in the queue provides the most useful information in determining how many vehicles are presently in the queue. For example, in Figure 1, three probes are waiting in the queue where the last probe is the 6$^{th}$ vehicle in the queue, the queue position of which is denoted by $L_p$. By utilizing $L_p$, the total number of probes in the queue, and the time data on which they join the queue, it might then be possible to estimate $\lambda$ and $p$. For example, $p$ can be estimated by dividing the number of probes in the queue by $L_p$. This method for estimating both $\lambda$ and $p$ is explained in detail after the next two sections.

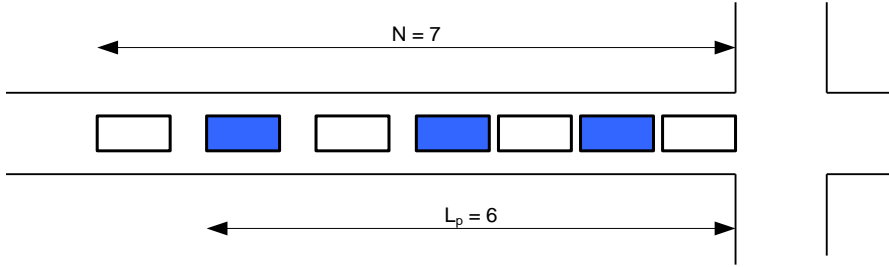

**Figure 1 Snapshot of an intersection right before the red interval terminates. Shaded boxes represent probe vehicles.**

**ESTIMATING UNKNOWN ARRIVAL RATE**

When the percent of probe vehicles in the traffic stream is known, the probe counts ($N_p$) can be expanded to estimate the total traffic flow. A relevant question is how much data need to be collected to reliably predict the unknown flow rate. The estimator to be used for the arrival rate $\lambda$ is shown above in Equation 3. Before investigating the data requirement, it is important to note that this estimator is unbiased since,

$$E[\hat{\lambda}] = E\left[\frac{N_p}{\Delta p}\right] = \frac{E[N_p]}{\Delta p} = \frac{\Delta \lambda p}{\Delta p} = \lambda \, . \qquad (5)$$

The variance of this estimator can also be obtained easily since $N_p$ is a Poisson process as explained in the previous section.

$$Var[\hat{\lambda}] = Var\left[\frac{N_p}{\Delta p}\right] = \frac{Var[N_p]}{(\Delta p)^2} = \frac{\Delta \lambda p}{(\Delta p)^2} = \frac{\lambda}{\Delta p} \qquad (6)$$

This equation shows that the variance will decrease as the data collection interval $\Delta$ and $p$ get larger. To determine how large these parameters need to be in order to estimate $\lambda$ within a given confidence level, the probability of the estimator $\hat{\lambda}$ falling in the confidence interval needs to be calculated. This can be written as follows for a given acceptable percent deviation ($\delta$) from the mean.

$$P\big((1-\delta)\lambda < \hat{\lambda} < (1+\delta)\lambda\big) = P\left((1-\delta)\lambda < \frac{N_p}{\Delta p} < (1+\delta)\lambda\right)$$
$$= P\big((1-\delta)\lambda\Delta p < N_p < (1+\delta)\lambda\Delta p\big) \qquad (7)$$

Since $Np$ is Poisson, the probability in Equation 7 can be readily calculated and compared to $\alpha$-levels to determine whether the flow rate can be estimated reliably for selected error criteria.

Three examples are created to analyze the sample size requirements as $\Delta$, $p$ and $\lambda$ vary. The results are summarized in Tables 1-3. Three different traffic flow level are considered: 400 vph (vehicles per hour), 800 vph, and 1200 vph. The precision level $\delta$ is selected to be 15% while $\alpha$ is assumed to be 10%. In other words, the requirement is to estimate the true $\lambda$ 90% of the time with a ± accuracy of 15%. The value in each cell of these tables is the probability of the estimator to be outside this 15% ± accuracy range. In other words, this value is one minus the probability calculated by Equation 7. If this probability is less than the 10% $\alpha$-level then, the corresponding cell is highlighted to indicate that the probe observations will be sufficient for estimating the $\lambda$ at the given confidence level. As it can be observed, highlighted cells are accumulated in the lower right corner in each table since both $p$ and $\Delta$ are larger there. Also, as flow rate increases the number of these highlighted cells increases as expected. It should also be noted that for smaller $\Delta$ periods (e.g., less than 15-minutes in Table 1) the targeted reliability can not be achieved even if all vehicles are probes.

## Approximation by Normal Distribution

The probability distribution of $\hat{\lambda}$ can be approximated with Normal distribution reasonably well especially when the number of data points (number of probes) is large. The mean and variance of $\hat{\lambda}$ are shown before in Equations 5 and 6. Then,

$$\hat{\lambda} \sim N(\lambda, \lambda/(\Delta p)). \qquad (8)$$

Using the Normal distribution, the probability of $\hat{\lambda}$ being within the confidence interval can be rewritten as follows:

$$P\big((1-\delta)\lambda < \hat{\lambda} < (1+\delta)\lambda\big) = 1 - 2F((1-\delta)\lambda) = 1 - 2\Phi\left(\frac{-\delta\lambda}{\sqrt{\lambda/\Delta p}}\right) \qquad (9)$$

where $F(.)$ is the Normal cumulative distribution function for $\hat{\lambda}$ and $\Phi(.)$ is the cumulative distribution for the standard Normal. The values in Table 1 are recreated as shown in Table 4 using the Normal distribution as opposed to Poisson. Table 4 conveys the same information as Table 1 except the probability values are slightly different. The difference between the values in two tables diminishes as $p$ and $\Delta$ increases. This shows that the Normal approximation works well. The same tables are created for $\lambda=800$ vph and $\lambda=1200$ vph flow rates but are not included in the paper since their correspondence with the respective exact solutions (given Tables 2 and 3) is even better since arrival rates are larger.

The results presented in Tables 1 to 4 are for specific flow rate values. They provide insights into how estimation error behaves at different data collection intervals and probe percentage levels. In terms of decision making, these types of results help determine what probe percentage is necessary to estimate flow rate reliably at different flow rate levels. In essence, the flow rate (which cold be an approximate value) is required as an input into this process. Therefore, to determine an appropriate $\Delta$ for field data collection/applications an iterative process is needed unless an approximate flow rate is known. The required data to be collected can be expressed in an alternative way to better facilitate the field applications. This alternative method does not require flow rate as an input as explained next.

The expected number of probe vehicles to be observed in $\Delta$ can be written as:
$$E[N_p] = \lambda \Delta p . \tag{10}$$
If we want to set the probability in Equation 9 to be equal to $1-\alpha$ then,
$$\Phi\left(\frac{-\delta\lambda}{\sqrt{\lambda/\Delta p}}\right) = \alpha/2 . \tag{11}$$
Or equivalently,
$$\frac{-\delta\lambda}{\sqrt{\lambda/\Delta p}} = Z_{\alpha/2} . \tag{12}$$

Equation 12 can be solved for $\lambda$ and substituted into Equation 10 to obtain the following expression,
$$E[N_p] = \lambda \Delta p = \frac{z_{\alpha/2}^2}{\delta^2} . \tag{13}$$
In other words, the expected number of probe observations in an interval should be equal or larger than the quantity shown in Equation 13. This quantity can be calculated without $\lambda$, $\Delta$ and $p$. In field applications, observation intervals yielding probe observations equal to or more than this quantity can be used to estimate the flow rate which will produce results that are reliable as specified by $\alpha$ and $\delta$. Figure 2 shows how $E[Np]$ changes with $\delta$ for the three selected $\alpha$-levels. For the analyses presented in Tables 1-4, $\delta$ is selected to be 15% and $\alpha$-level is assumed to be 10%. This corresponds to 120 probe observations. It can be verified that all the highlighted cells in Tables 1-4 correspond to expected probe observations (which can be calculated from Equation 10) that are more than this threshold value (i.e., 120). It can be seen from Figure 2 that decreasing $\delta$ beyond 10% or 15% dramatically increases the probe requirements.

**TABLE 1 Probability of the estimator to be outside of the upper or lower thresholds when $\lambda$ = 400 vph**

| Δ | Percentage of Probe Vehicles | | | | | | | | | | | | | | | | | | |
|---|---|---|---|---|---|---|---|---|---|---|---|---|---|---|---|---|---|---|---|
|   | 5% | 6% | 7% | 8% | 9% | 10% | 15% | 20% | 25% | 30% | 35% | 40% | 45% | 50% | 60% | 70% | 80% | 90% | 100% |
| 1 | 1.00 | 1.00 | 1.00 | 1.00 | 1.00 | 1.00 | 0.63 | 1.00 | 1.00 | 0.73 | 0.74 | 0.78 | 0.78 | 0.78 | 0.80 | 0.64 | 0.67 | 0.84 | 0.70 |
| 2 | 1.00 | 1.00 | 0.63 | 0.63 | 1.00 | 1.00 | 0.73 | 0.78 | 0.78 | 0.80 | 0.64 | 0.67 | 0.84 | 0.70 | 0.60 | 0.62 | 0.65 | 0.67 | 0.59 |
| 5 | 1.00 | 0.73 | 0.74 | 0.78 | 0.78 | 0.78 | 0.82 | 0.70 | 0.73 | 0.64 | 0.56 | 0.59 | 0.52 | 0.55 | 0.51 | 0.47 | 0.44 | 0.41 | 0.39 |
| 10 | 0.78 | 0.80 | 0.64 | 0.67 | 0.84 | 0.70 | 0.64 | 0.59 | 0.55 | 0.51 | 0.47 | 0.44 | 0.41 | 0.39 | 0.35 | 0.30 | 0.27 | 0.25 | 0.22 |
| 15 | 0.82 | 0.84 | 0.57 | 0.60 | 0.62 | 0.64 | 0.52 | 0.51 | 0.48 | 0.41 | 0.35 | 0.35 | 0.33 | 0.29 | 0.25 | 0.21 | 0.18 | 0.15 | 0.13 |
| 20 | 0.70 | 0.60 | 0.62 | 0.65 | 0.67 | 0.59 | 0.51 | 0.44 | 0.39 | 0.35 | 0.30 | 0.27 | 0.25 | 0.22 | 0.18 | 0.15 | 0.12 | 0.10 | 0.08 |
| 30 | 0.64 | 0.67 | 0.50 | 0.53 | 0.56 | 0.51 | 0.41 | 0.35 | 0.29 | 0.25 | 0.21 | 0.18 | 0.15 | 0.13 | 0.10 | 0.08 | 0.06 | 0.04 | 0.03 |
| 60 | 0.51 | 0.47 | 0.39 | 0.43 | 0.36 | 0.35 | 0.25 | 0.18 | 0.13 | 0.10 | 0.08 | 0.06 | 0.04 | 0.03 | 0.02 | 0.01 | 0.01 | 0.00 | 0.00 |
| 120 | 0.35 | 0.28 | 0.26 | 0.23 | 0.22 | 0.18 | 0.10 | 0.06 | 0.03 | 0.02 | 0.01 | 0.01 | 0.00 | 0.00 | 0.00 | 0.00 | 0.00 | 0.00 | 0.00 |

**TABLE 2 Probability of the estimator to be outside of the upper or lower thresholds when $\lambda$ = 800 vph**

| Δ | Percentage of Probe Vehicles | | | | | | | | | | | | | | | | | | |
|---|---|---|---|---|---|---|---|---|---|---|---|---|---|---|---|---|---|---|---|
|   | 5% | 6% | 7% | 8% | 9% | 10% | 15% | 20% | 25% | 30% | 35% | 40% | 45% | 50% | 60% | 70% | 80% | 90% | 100% |
| 1 | 1.00 | 1.00 | 0.63 | 0.63 | 1.00 | 1.00 | 0.73 | 0.78 | 0.78 | 0.80 | 0.64 | 0.67 | 0.84 | 0.70 | 0.60 | 0.62 | 0.65 | 0.67 | 0.59 |
| 2 | 1.00 | 1.00 | 0.73 | 0.73 | 1.00 | 0.78 | 0.80 | 0.67 | 0.70 | 0.60 | 0.62 | 0.65 | 0.67 | 0.59 | 0.53 | 0.49 | 0.52 | 0.47 | 0.44 |
| 5 | 0.78 | 0.80 | 0.64 | 0.67 | 0.84 | 0.70 | 0.64 | 0.59 | 0.55 | 0.51 | 0.47 | 0.44 | 0.41 | 0.39 | 0.35 | 0.30 | 0.27 | 0.25 | 0.22 |
| 10 | 0.70 | 0.60 | 0.62 | 0.65 | 0.67 | 0.59 | 0.51 | 0.44 | 0.39 | 0.35 | 0.30 | 0.27 | 0.25 | 0.22 | 0.18 | 0.15 | 0.12 | 0.10 | 0.08 |
| 15 | 0.64 | 0.67 | 0.50 | 0.53 | 0.56 | 0.51 | 0.41 | 0.35 | 0.29 | 0.25 | 0.21 | 0.18 | 0.15 | 0.13 | 0.10 | 0.08 | 0.06 | 0.04 | 0.03 |
| 20 | 0.59 | 0.53 | 0.49 | 0.52 | 0.47 | 0.44 | 0.35 | 0.27 | 0.22 | 0.18 | 0.15 | 0.12 | 0.10 | 0.08 | 0.06 | 0.04 | 0.03 | 0.02 | 0.01 |
| 30 | 0.51 | 0.47 | 0.39 | 0.43 | 0.36 | 0.35 | 0.25 | 0.18 | 0.13 | 0.10 | 0.08 | 0.06 | 0.04 | 0.03 | 0.02 | 0.01 | 0.01 | 0.00 | 0.00 |
| 60 | 0.35 | 0.28 | 0.26 | 0.23 | 0.22 | 0.18 | 0.10 | 0.06 | 0.03 | 0.02 | 0.01 | 0.01 | 0.00 | 0.00 | 0.00 | 0.00 | 0.00 | 0.00 | 0.00 |
| 120 | 0.18 | 0.14 | 0.12 | 0.08 | 0.07 | 0.06 | 0.02 | 0.01 | 0.00 | 0.00 | 0.00 | 0.00 | 0.00 | 0.00 | 0.00 | 0.00 | 0.00 | 0.00 | 0.00 |

**TABLE 3 Probability of the estimator to be outside of the upper or lower thresholds when $\lambda$ = 1200 vph**

| Δ | Percentage of Probe Vehicles | | | | | | | | | | | | | | | | | | |
|---|---|---|---|---|---|---|---|---|---|---|---|---|---|---|---|---|---|---|---|
|   | 5% | 6% | 7% | 8% | 9% | 10% | 15% | 20% | 25% | 30% | 35% | 40% | 45% | 50% | 60% | 70% | 80% | 90% | 100% |
| 1 | 0.63 | 1.00 | 1.00 | 1.00 | 0.73 | 0.73 | 0.78 | 0.80 | 0.82 | 0.84 | 0.57 | 0.60 | 0.62 | 0.64 | 0.67 | 0.50 | 0.53 | 0.56 | 0.51 |
| 2 | 0.73 | 1.00 | 0.78 | 0.78 | 0.81 | 0.80 | 0.84 | 0.60 | 0.64 | 0.67 | 0.50 | 0.53 | 0.56 | 0.51 | 0.47 | 0.39 | 0.43 | 0.36 | 0.35 |
| 5 | 0.82 | 0.84 | 0.57 | 0.60 | 0.62 | 0.64 | 0.52 | 0.51 | 0.48 | 0.41 | 0.35 | 0.35 | 0.33 | 0.29 | 0.25 | 0.21 | 0.18 | 0.15 | 0.13 |
| 10 | 0.64 | 0.67 | 0.50 | 0.53 | 0.56 | 0.51 | 0.41 | 0.35 | 0.29 | 0.25 | 0.21 | 0.18 | 0.15 | 0.13 | 0.10 | 0.08 | 0.06 | 0.04 | 0.03 |
| 15 | 0.52 | 0.56 | 0.44 | 0.47 | 0.39 | 0.41 | 0.33 | 0.25 | 0.18 | 0.15 | 0.13 | 0.10 | 0.08 | 0.07 | 0.04 | 0.03 | 0.02 | 0.01 | 0.01 |
| 20 | 0.51 | 0.47 | 0.39 | 0.43 | 0.36 | 0.35 | 0.25 | 0.18 | 0.13 | 0.10 | 0.08 | 0.06 | 0.04 | 0.03 | 0.02 | 0.01 | 0.01 | 0.00 | 0.00 |
| 30 | 0.41 | 0.36 | 0.32 | 0.28 | 0.25 | 0.25 | 0.15 | 0.10 | 0.07 | 0.04 | 0.03 | 0.02 | 0.01 | 0.01 | 0.00 | 0.00 | 0.00 | 0.00 | 0.00 |
| 60 | 0.25 | 0.22 | 0.17 | 0.14 | 0.11 | 0.10 | 0.04 | 0.02 | 0.01 | 0.00 | 0.00 | 0.00 | 0.00 | 0.00 | 0.00 | 0.00 | 0.00 | 0.00 | 0.00 |
| 120 | 0.10 | 0.07 | 0.05 | 0.04 | 0.03 | 0.02 | 0.00 | 0.00 | 0.00 | 0.00 | 0.00 | 0.00 | 0.00 | 0.00 | 0.00 | 0.00 | 0.00 | 0.00 | 0.00 |

**TABLE 4 Probability of the estimator to be outside of the upper or lower thresholds when $\lambda$ = 400 vph (calculated by Normal approximation)**

| | Percentage of Probe Vehicles | | | | | | | | | | | | | | | | | | |
|---|---|---|---|---|---|---|---|---|---|---|---|---|---|---|---|---|---|---|---|
| $\Delta$ | 5% | 6% | 7% | 8% | 9% | 10% | 15% | 20% | 25% | 30% | 35% | 40% | 45% | 50% | 60% | 70% | 80% | 90% | 100% |
| 1 | 0.93 | 0.92 | 0.92 | 0.91 | 0.91 | 0.90 | 0.88 | 0.86 | 0.85 | 0.83 | 0.82 | 0.81 | 0.80 | 0.78 | 0.76 | 0.75 | 0.73 | 0.71 | 0.70 |
| 2 | 0.90 | 0.89 | 0.88 | 0.88 | 0.87 | 0.86 | 0.83 | 0.81 | 0.78 | 0.76 | 0.75 | 0.73 | 0.71 | 0.70 | 0.67 | 0.65 | 0.62 | 0.60 | 0.58 |
| 5 | 0.85 | 0.83 | 0.82 | 0.81 | 0.80 | 0.78 | 0.74 | 0.70 | 0.67 | 0.64 | 0.61 | 0.58 | 0.56 | 0.54 | 0.50 | 0.47 | 0.44 | 0.41 | 0.39 |
| 10 | 0.78 | 0.76 | 0.75 | 0.73 | 0.71 | 0.70 | 0.64 | 0.58 | 0.54 | 0.50 | 0.47 | 0.44 | 0.41 | 0.39 | 0.34 | 0.31 | 0.27 | 0.25 | 0.22 |
| 15 | 0.74 | 0.71 | 0.69 | 0.67 | 0.65 | 0.64 | 0.56 | 0.50 | 0.45 | 0.41 | 0.37 | 0.34 | 0.31 | 0.29 | 0.25 | 0.21 | 0.18 | 0.15 | 0.13 |
| 20 | 0.70 | 0.67 | 0.65 | 0.62 | 0.60 | 0.58 | 0.50 | 0.44 | 0.39 | 0.34 | 0.31 | 0.27 | 0.25 | 0.22 | 0.18 | 0.15 | 0.12 | 0.10 | 0.08 |
| 30 | 0.64 | 0.60 | 0.57 | 0.55 | 0.52 | 0.50 | 0.41 | 0.34 | 0.29 | 0.25 | 0.21 | 0.18 | 0.15 | 0.13 | 0.10 | 0.08 | 0.06 | 0.04 | 0.03 |
| 60 | 0.50 | 0.46 | 0.43 | 0.40 | 0.37 | 0.34 | 0.25 | 0.18 | 0.13 | 0.10 | 0.08 | 0.06 | 0.04 | 0.03 | 0.02 | 0.01 | 0.01 | 0.00 | 0.00 |
| 120 | 0.34 | 0.30 | 0.26 | 0.23 | 0.20 | 0.18 | 0.10 | 0.06 | 0.03 | 0.02 | 0.01 | 0.01 | 0.00 | 0.00 | 0.00 | 0.00 | 0.00 | 0.00 | 0.00 |

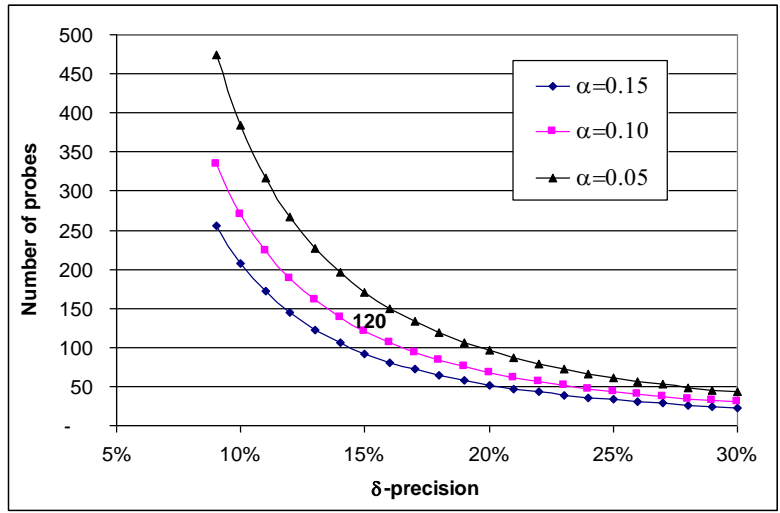

**Figure 2 Required number of probe observations versus precision or acceptable percent deviation from the true mean at three different confidence levels**

## ESTIMATING UNKNOWN PROBE PERCENTAGE

When the flow rate $\lambda$ is known but the percent of probe vehicles $p$ in the traffic stream is not known, the probe counts ($N_p$) can be as indicated in Equation 4 to estimate $p$. As before, it can be verified that the estimator of $p$ is also unbiased:

$$E[\hat{p}] = E\left[\frac{N_p}{\Delta \lambda}\right] = \frac{E[N_p]}{\Delta \lambda} = \frac{\Delta \lambda p}{\Delta \lambda} = p. \tag{14}$$

The variance of this estimator can also be obtained easily since $N_p$ is a Poisson process as explained in the previous section.

$$Var[\hat{p}] = Var\left[\frac{N_p}{\Delta \lambda}\right] = \frac{Var[N_p]}{(\Delta \lambda)^2} = \frac{\Delta \lambda p}{(\Delta \lambda)^2} = \frac{p}{\Delta \lambda} \tag{15}$$

This equation shows that the variance will decrease as the data collection interval $\Delta$ and $\lambda$ get larger. To determine how large these parameters need to be in order to estimate $p$ within a given confidence level, the probability of the estimator $\hat{p}$ falling in the confidence interval needs to calculated. This can be written as follows for a given acceptable percent deviation ($\delta$) from the mean.

$$P((1-\delta)p < \hat{p} < (1+\delta)p) = P\left((1-\delta)\lambda < \frac{N_p}{\Delta \lambda} < (1+\delta)\lambda\right)$$
$$= P((1-\delta)\lambda \Delta p < N_p < (1+\delta)\lambda \Delta p) \tag{16}$$

This, due to the symmetry in the two estimator $\hat{p}$ and $\hat{\lambda}$, turns out to be identical to the expression in Equation 7. Therefore, the required sample size to estimate either $\lambda$ or $p$ at a given confidence level will be the same. In other words, Tables 1-4 and Figure 2 are equally applicable to both $\hat{p}$ and $\hat{\lambda}$.

**ESTIMATING BOTH ARRIVAL RATE AND PROBE PERCENTAGE**

As explained in Methodology section, probe counts in an interval $\Delta$ alone are not sufficient to estimate both the arrival or flow rate $\lambda$ and the market penetration rate $p$ simultaneously. Therefore, the location of the last probe vehicle in a queue $L_p$ at a signal (see Figure 1) and the time it joins the back of the queue (denoted by $T_p$ and measured in reference to the beginning of red phase) will be used as additional information. Since $Lp$ gives the total number of all vehicles (including the last probe itself) ahead of the last probe and the total probes arrived on red can be counted (denoted by $Np$), one potential method to estimate $p$ is as follows:

$$\hat{p}_1 = \frac{N_p}{L_p}. \tag{17}$$

Since the arrival times of probes are also collected, the total time for $Lp$ arrivals is simply equal to the time the last probe joins the back of the queue, $Tp$. An arrival rate can be calculated based on these two parameters to potentially estimate the $\lambda$:

$$\hat{\lambda}_1 = \frac{L_p}{T_p}. \tag{18}$$

The expected value and variance of these estimators can not be easily calculated as it was done for the estimators in the previous sections since these depend on two random variables with relatively more complex pdfs (*8*). Therefore, simulation is used in this section to perform the analyses. A custom program in C++ is developed to simulate operations at an intersection approach where vehicles are assumed to queue vertically and overflow queue (or the residual queue at the end of green period) is ignored. Vehicle arrivals are Poisson as before. For the purpose of this study, these assumptions are not critical. Cycle length is assumed to be 120 seconds and red phase 60 seconds. The objective is to estimate both $\lambda$ and $p$ accurately and assess the sample size needed for reliable estimation. For brevity, the analyses are done only for one arrival rate where $\lambda = 1200$ vph or 0.333 vehicle per second.

The simulation model is first run to generate data to analyze the expected value and variance of the estimators shown in Equations 17 and 18. The results of ten million replicas (the duration of each one is equal to the red duration) are shown in Table 5 where the first two rows provide the results for $\lambda$ and next two rows for $p$ as labeled. (The last two row of this table are explained below.) The true value of $\lambda$ is 0.333. However, as it can be observed in the first row, the expected value fluctuates and does not converge to this true value even at high probe levels. Also, the variance is very large, especially at small $p$ levels. For $p$, the true values are indicated as percentages at the top of the table. The expected value of the estimator in Equation 17 consistently yields biased results as can be observed in the table. For example, at 20% probe level the expected value of $p$ turns out to be 25.8%. Obviously, the estimators of Equations 17 and 18 do not seem to provide reliable results from this initial investigation.

**TABLE 5 Performance of estimators at different probe percentage levels $p$**

| | $p$ | 5% | 10% | 15% | 20% | 25% | 30% | 35% | 40% | 45% | 50% | 60% | 70% | 80% | 90% | 100% |
|---|---|---|---|---|---|---|---|---|---|---|---|---|---|---|---|---|
| Method 1 for $\lambda$ | E(Lp/Tp) | 0.306 | 0.371 | 0.373 | 0.365 | 0.362 | 0.357 | 0.354 | 0.353 | 0.353 | 0.352 | 0.352 | 0.351 | 0.351 | 0.351 | 0.351 |
| | VAR(Lp/Tp) | 119.479 | 99.283 | 61.940 | 15.451 | 26.833 | 3.294 | 0.267 | 0.018 | 0.025 | 0.019 | 0.006 | 0.006 | 0.006 | 0.006 | 0.006 |
| Method 1 for $p$ | E(Np/Lp) | 0.116 | 0.174 | 0.217 | 0.258 | 0.300 | 0.344 | 0.390 | 0.435 | 0.482 | 0.528 | 0.622 | 0.716 | 0.811 | 0.905 | 1.000 |
| | VAR(Np/Lp) | 0.029 | 0.025 | 0.019 | 0.016 | 0.014 | 0.014 | 0.013 | 0.014 | 0.014 | 0.013 | 0.012 | 0.011 | 0.008 | 0.005 | 0.000 |
| Method 2 for $p$ | E($p_2$) | 0.071 | 0.119 | 0.163 | 0.208 | 0.256 | 0.305 | 0.354 | 0.403 | 0.452 | 0.502 | 0.602 | 0.701 | 0.801 | 0.900 | 1.000 |
| | VAR($p_2$) | 0.021 | 0.019 | 0.015 | 0.013 | 0.012 | 0.013 | 0.013 | 0.014 | 0.014 | 0.014 | 0.014 | 0.012 | 0.009 | 0.005 | 0.000 |

The estimator for $p$ in Equation 17 gives consistently larger predictions because in every red interval there are Np probes but there are more than $Lp$ vehicles in the queue. In other words, the denominator is being underestimated. To correct for that, the following alternative estimator is proposed.

$$\hat{p}_2 = \frac{N_p}{L_p + (R - T_p)(L_p - N_p)/T_p}. \tag{19}$$

This estimator, accounts for the arrivals beyond the last probe by calculating a rate for non-probe arrivals and multiplying this calculated rate with the duration after the arrival of Lp beyond which no probe is observed. The performance of this estimator is also included in Table 5. The last two rows provide the expected value and variance of this estimator of $p$. Compared to the previous estimator; this gives more accurate estimates with less variance. For example, at 20% probe level the expected value of $p$ is now 20.8%. Therefore, this estimator is used to estimate $p$. The results of Table 5 are for one cycle ($\Delta = 2$ minutes). In other words, the $p$ is estimated based on data collected only during a single cycle. As the data collection period $\Delta$ increases, the expected value of $p$ does not change but the variance decreases since more samples are used in the estimation. For example, if the data is collected in two cycles and $p$ is estimated at the end of the second cycle then, the variance will be half of what is shown in Table 5.

In order to see how much data need to be collected to estimate $p$ reliably, Table 6 is created where each row represents a different data collection period in minutes. Similar to the tables in the previous section (Tables 1-4), the precision level $\delta$ is selected to be 15% while $\alpha$ is assumed to be 10%. In other words, the requirement is to estimate the true $p$ 90% of the time with a ± accuracy of 15%. The value in each cell of these tables is the probability of the estimator to be outside this 15% ± accuracy range. If this probability is less than the $\alpha$-level (i.e., 0.10) then, the corresponding cell is highlighted to indicate that the probe observations will be sufficient for estimating the $p$ at the given confidence level. As it can be observed, at large $p$ levels the estimates are more reliable. For lower than 25% probe levels, more than one-hour worth of data needs to be collected to be able to estimate $p$ reliably.

**TABLE 6 Probability of the estimated *p* to be outside of the upper or lower thresholds (calculated by Normal approximation)**

| | Percentage of Probe Vehicles | | | | | | | | | | | | | | | | | | |
|---|---|---|---|---|---|---|---|---|---|---|---|---|---|---|---|---|---|---|---|
| Δ | 5% | 6% | 7% | 8% | 9% | 10% | 15% | 20% | 25% | 30% | 35% | 40% | 45% | 50% | 60% | 70% | 80% | 90% | 100% |
| 2 | 0.96 | 0.95 | 0.94 | 0.93 | 0.92 | 0.91 | 0.85 | 0.79 | 0.74 | 0.69 | 0.65 | 0.61 | 0.57 | 0.53 | 0.44 | 0.33 | 0.20 | 0.06 | 0.00 |
| 4 | 0.94 | 0.93 | 0.92 | 0.91 | 0.89 | 0.88 | 0.80 | 0.71 | 0.64 | 0.57 | 0.52 | 0.47 | 0.42 | 0.37 | 0.27 | 0.17 | 0.07 | 0.01 | 0.00 |
| 6 | 0.93 | 0.92 | 0.90 | 0.89 | 0.87 | 0.85 | 0.75 | 0.65 | 0.56 | 0.49 | 0.43 | 0.37 | 0.32 | 0.28 | 0.18 | 0.09 | 0.03 | 0.00 | 0.00 |
| 8 | 0.92 | 0.91 | 0.89 | 0.87 | 0.85 | 0.83 | 0.72 | 0.60 | 0.50 | 0.43 | 0.36 | 0.31 | 0.25 | 0.21 | 0.12 | 0.05 | 0.01 | 0.00 | 0.00 |
| 10 | 0.91 | 0.90 | 0.88 | 0.86 | 0.84 | 0.81 | 0.69 | 0.56 | 0.46 | 0.37 | 0.31 | 0.25 | 0.20 | 0.16 | 0.08 | 0.03 | 0.00 | 0.00 | 0.00 |
| 20 | 0.88 | 0.86 | 0.84 | 0.81 | 0.78 | 0.75 | 0.58 | 0.42 | 0.29 | 0.21 | 0.15 | 0.10 | 0.07 | 0.05 | 0.01 | 0.00 | 0.00 | 0.00 | 0.00 |
| 40 | 0.85 | 0.82 | 0.79 | 0.76 | 0.72 | 0.68 | 0.46 | 0.27 | 0.15 | 0.08 | 0.04 | 0.02 | 0.01 | 0.00 | 0.00 | 0.00 | 0.00 | 0.00 | 0.00 |
| 60 | 0.84 | 0.81 | 0.77 | 0.73 | 0.69 | 0.65 | 0.39 | 0.18 | 0.08 | 0.03 | 0.01 | 0.01 | 0.00 | 0.00 | 0.00 | 0.00 | 0.00 | 0.00 | 0.00 |

As mentioned above, the estimator for λ given in Equation 18 does not perform very well. An alternative method is to use the same estimator of Equation 3 with a modification as shown below.

$$\hat{\lambda} = \frac{N_p}{\Delta \hat{p}_2} \quad (20)$$

The total number of probe counts observed in Δ (during both green and red intervals) is divided by the estimated *p* to estimate the unknown λ. Table 7 and 8 presents the expected value and variance of this new estimator for λ obtained based on the data generated from the simulation runs. This estimator performs much better since variance is smaller and the estimates are unbiased. The variance approaches zero as Δ and *p* increase.

Similar to the previous tables (Tables 1-4 and 6) Table 9 is constructed to analyze how the reliability of the estimated λ is changing by Δ and *p*. Since this table is based on 1200 vph arrival rate it should be compared and contrasted with Table 3 which contains results for the same rate. Overall the results of Table 3 are better since the corresponding cell values are smaller in Table 3. This is expected since Table 3 values are based on known true p levels.

**TABLE 7 Expected value of the estimator for λ given by Equation 20 (True λ is 0.333)**

| | Percentage of Probe Vehicles | | | | | | | | | | | | | | | | | | |
|---|---|---|---|---|---|---|---|---|---|---|---|---|---|---|---|---|---|---|---|
| Δ | 5% | 6% | 7% | 8% | 9% | 10% | 15% | 20% | 25% | 30% | 35% | 40% | 45% | 50% | 60% | 70% | 80% | 90% | 100% |
| 2 | 0.19 | 0.22 | 0.25 | 0.27 | 0.29 | 0.31 | 0.36 | 0.37 | 0.37 | 0.36 | 0.36 | 0.35 | 0.35 | 0.34 | 0.34 | 0.34 | 0.34 | 0.33 | 0.33 |
| 4 | 0.31 | 0.33 | 0.35 | 0.36 | 0.36 | 0.37 | 0.36 | 0.35 | 0.35 | 0.34 | 0.34 | 0.34 | 0.34 | 0.34 | 0.34 | 0.34 | 0.34 | 0.33 | 0.33 |
| 6 | 0.35 | 0.36 | 0.36 | 0.36 | 0.36 | 0.35 | 0.34 | 0.34 | 0.34 | 0.34 | 0.34 | 0.34 | 0.34 | 0.34 | 0.33 | 0.33 | 0.33 | 0.33 | 0.33 |
| 8 | 0.36 | 0.35 | 0.35 | 0.34 | 0.34 | 0.34 | 0.33 | 0.33 | 0.34 | 0.34 | 0.34 | 0.34 | 0.33 | 0.33 | 0.33 | 0.33 | 0.33 | 0.33 | 0.33 |
| 10 | 0.35 | 0.34 | 0.33 | 0.33 | 0.33 | 0.33 | 0.33 | 0.33 | 0.33 | 0.33 | 0.33 | 0.33 | 0.33 | 0.33 | 0.33 | 0.33 | 0.33 | 0.33 | 0.33 |
| 20 | 0.30 | 0.30 | 0.30 | 0.30 | 0.30 | 0.31 | 0.32 | 0.33 | 0.33 | 0.33 | 0.33 | 0.33 | 0.33 | 0.33 | 0.33 | 0.33 | 0.33 | 0.33 | 0.33 |
| 40 | 0.27 | 0.28 | 0.28 | 0.29 | 0.29 | 0.30 | 0.31 | 0.32 | 0.33 | 0.33 | 0.33 | 0.33 | 0.33 | 0.33 | 0.33 | 0.33 | 0.33 | 0.33 | 0.33 |
| 60 | 0.26 | 0.27 | 0.27 | 0.28 | 0.29 | 0.29 | 0.31 | 0.32 | 0.33 | 0.33 | 0.33 | 0.33 | 0.33 | 0.33 | 0.33 | 0.33 | 0.33 | 0.33 | 0.33 |

**TABLE 8 Variance for the estimator of λ given by Equation 20**

| | Percentage of Probe Vehicles | | | | | | | | | | | | | | | | | | |
|---|---|---|---|---|---|---|---|---|---|---|---|---|---|---|---|---|---|---|---|
| Δ | 5% | 6% | 7% | 8% | 9% | 10% | 15% | 20% | 25% | 30% | 35% | 40% | 45% | 50% | 60% | 70% | 80% | 90% | 100% |
| 2 | 0.051 | 0.056 | 0.062 | 0.064 | 0.067 | 0.071 | 0.061 | 0.049 | 0.035 | 0.025 | 0.018 | 0.013 | 0.010 | 0.008 | 0.006 | 0.004 | 0.004 | 0.003 | 0.003 |
| 4 | 0.078 | 0.119 | 0.076 | 0.075 | 0.067 | 0.062 | 0.037 | 0.019 | 0.011 | 0.008 | 0.006 | 0.005 | 0.004 | 0.003 | 0.003 | 0.002 | 0.002 | 0.002 | 0.001 |
| 6 | 0.101 | 0.074 | 0.064 | 0.052 | 0.046 | 0.039 | 0.016 | 0.009 | 0.006 | 0.004 | 0.004 | 0.003 | 0.003 | 0.002 | 0.002 | 0.001 | 0.001 | 0.001 | 0.001 |
| 8 | 0.072 | 0.060 | 0.047 | 0.038 | 0.032 | 0.024 | 0.010 | 0.006 | 0.004 | 0.003 | 0.003 | 0.002 | 0.002 | 0.002 | 0.001 | 0.001 | 0.001 | 0.001 | 0.001 |
| 10 | 0.064 | 0.045 | 0.035 | 0.026 | 0.020 | 0.016 | 0.008 | 0.005 | 0.003 | 0.002 | 0.002 | 0.002 | 0.001 | 0.001 | 0.001 | 0.001 | 0.001 | 0.001 | 0.001 |
| 20 | 0.021 | 0.016 | 0.012 | 0.010 | 0.009 | 0.007 | 0.004 | 0.002 | 0.002 | 0.001 | 0.001 | 0.001 | 0.001 | 0.001 | 0.001 | 0.000 | 0.000 | 0.000 | 0.000 |
| 40 | 0.010 | 0.008 | 0.007 | 0.006 | 0.005 | 0.004 | 0.002 | 0.001 | 0.001 | 0.001 | 0.000 | 0.000 | 0.000 | 0.000 | 0.000 | 0.000 | 0.000 | 0.000 | 0.000 |
| 60 | 0.007 | 0.006 | 0.005 | 0.004 | 0.003 | 0.003 | 0.001 | 0.001 | 0.001 | 0.000 | 0.000 | 0.000 | 0.000 | 0.000 | 0.000 | 0.000 | 0.000 | 0.000 | 0.000 |

**TABLE 9 Probability of the estimated λ to be outside of the upper or lower thresholds (calculated by Normal approximation)**

| Δ | \multicolumn{19}{c|}{Percentage of Probe Vehicles} |
|---|---|---|---|---|---|---|---|---|---|---|---|---|---|---|---|---|---|---|---|
|   | 5% | 6% | 7% | 8% | 9% | 10% | 15% | 20% | 25% | 30% | 35% | 40% | 45% | 50% | 60% | 70% | 80% | 90% | 100% |
| 2 | 0.85 | 0.85 | 0.86 | 0.86 | 0.87 | 0.87 | 0.90 | 0.91 | 0.92 | 0.93 | 0.93 | 0.94 | 0.94 | 0.94 | 0.95 | 0.95 | 0.96 | 0.96 | 0.96 |
| 4 | 0.83 | 0.83 | 0.84 | 0.84 | 0.85 | 0.85 | 0.84 | 0.82 | 0.79 | 0.75 | 0.71 | 0.66 | 0.62 | 0.58 | 0.51 | 0.46 | 0.41 | 0.38 | 0.34 |
| 6 | 0.86 | 0.89 | 0.86 | 0.86 | 0.85 | 0.84 | 0.79 | 0.72 | 0.63 | 0.56 | 0.51 | 0.47 | 0.43 | 0.39 | 0.33 | 0.28 | 0.24 | 0.21 | 0.18 |
| 8 | 0.88 | 0.85 | 0.84 | 0.83 | 0.82 | 0.80 | 0.69 | 0.60 | 0.51 | 0.45 | 0.40 | 0.35 | 0.32 | 0.28 | 0.23 | 0.19 | 0.15 | 0.12 | 0.10 |
| 10 | 0.85 | 0.84 | 0.82 | 0.80 | 0.78 | 0.75 | 0.62 | 0.52 | 0.44 | 0.37 | 0.32 | 0.28 | 0.24 | 0.22 | 0.16 | 0.13 | 0.10 | 0.08 | 0.06 |
| 20 | 0.84 | 0.82 | 0.79 | 0.76 | 0.73 | 0.70 | 0.57 | 0.47 | 0.38 | 0.32 | 0.26 | 0.22 | 0.19 | 0.16 | 0.12 | 0.09 | 0.06 | 0.05 | 0.03 |
| 40 | 0.75 | 0.72 | 0.69 | 0.66 | 0.63 | 0.60 | 0.44 | 0.31 | 0.21 | 0.15 | 0.11 | 0.08 | 0.06 | 0.05 | 0.03 | 0.02 | 0.01 | 0.00 | 0.00 |
| 60 | 0.70 | 0.67 | 0.64 | 0.60 | 0.57 | 0.53 | 0.33 | 0.17 | 0.09 | 0.04 | 0.02 | 0.01 | 0.01 | 0.00 | 0.00 | 0.00 | 0.00 | 0.00 | 0.00 |

**CONCLUSIONS**

This paper presents an investigation of flow rate estimation from probe vehicle data. Three cases are considered depending on the available information on the percentage of probes $p$ and the flow rate $\lambda$: i) $p$ is known but $\lambda$ is unknown ii) $\lambda$ is known but $p$ is unknown, and iii) both parameters are not known. Estimation methods for all three cases are presented along with the reliability of these estimates. For the first two cases, count data provide sufficient information to estimate the unknown parameters $p$ and $\lambda$ individually when the other one is known. For these two cases, simple analytical expressions are derived to analyze the accuracy and sample size requirements. However, when both parameters are unknown then additional information beyond count data is required. The position of probe vehicles in a queue at signalized intersections is used as additional information in that case. The results show for how long data need to be collected to estimate the parameters at acceptable confidence levels. When both $\lambda$ and $p$ are unknown data need to be collected over a longer period as compared to the first two cases to estimate the flow rate reliably.

The use of probe vehicle data has been investigated in previous studies for speed and travel time estimation where as low as 1% probe vehicles can provide reliable estimation. In comparison, the probe sample size requirements for flow rate estimation is much higher as shown in the analyses presented here for Poisson arrivals. Further research is needed to investigate other arrival types and estimation of time-dependent flow rate.